\documentclass[aps,prb,twocolumn]{revtex4}
\usepackage{graphicx,subfigure}
\usepackage{bm}
\usepackage{amsfonts}
\usepackage{dsfont}
\usepackage{color}
\begin{document}

\title{Entanglement spectrum of a topological phase in one dimension}
\author{Frank Pollmann}
\affiliation{Department of Physics, University of California, Berkeley CA 94720, USA}
\author{Erez Berg}
\affiliation{Department of Physics, Stanford University, Stanford, CA 94305, USA}
\affiliation{Department of Physics, Harvard University, Cambridge, MA 02138, USA}
\author{Ari M. Turner}
\affiliation{Department of Physics, University of California, Berkeley CA 94720, USA}
\date{\today}
\author{Masaki Oshikawa}
\affiliation{Institute for Solid State Physics, University of
Tokyo, Kashiwa 277-8581 Japan}

\begin{abstract}
We show that the Haldane phase of $S=1$ chains is characterized by
a double degeneracy of the entanglement spectrum. The degeneracy
is protected by a set of symmetries (either the dihedral group of
$\pi$-rotations about two orthogonal axes, time-reversal symmetry,
or bond centered inversion symmetry), and cannot be lifted unless
either a phase boundary to another, ``topologically trivial'',
phase is crossed, or the symmetry is broken. More generally, these
results offer a scheme to classify gapped phases of one
dimensional systems. Physically, the degeneracy of the
entanglement spectrum can be observed by adiabatically weakening a
bond to zero, which leaves the two disconnected halves of the
system in a finitely entangled state.
\end{abstract}

\maketitle

\section{Introduction}

A topological phase is a phase of matter which cannot be characterized by a
local order parameter, and thus falls beyond the Landau paradigm of
condensed matter physics. Topological phases are typically characterized by
a gap separating excitations from the ground state in the bulk and by the
presence of gapless edge modes.
The existence of edge excitations implies that a topological phase cannot be
deformed continuously into a conventional, topologically trivial phase
without going through a phase transition, in which the gap closes and the
edge mode merges with the bulk.

The Haldane phase of integer spin chains \cite{Haldane-1983,Haldane-1983a}
is an example of a \textquotedblleft symmetry protected topological
phase\textquotedblright\ in one dimension.\cite{Gu-2009} This phase appears
also in other one dimensional systems, such as chains of interacting bosons%
\cite{DallaTorre-2006} and fermions.\cite{Capponi-2009} These
gapped phases lack a local order parameter, and are not amenable
to a description by a site factorizable wave function.
Alternatively, in certain cases, the Haldane phase can be
characterized by the existence of fractionalized edge excitations,
by a non-vanishing non-local \textquotedblleft
string\textquotedblright\ order parameter or, as recently
proposed, by a quantized Berry
phase.\cite{denNijs-1989,Kennedy-1992,Hirano-2008}

However, in the most general case, the description of the Haldane
phase in terms of a string order parameter or spin-$\frac{1}{2}$
edge states is insufficient. As we will demonstrate below,
slightly deforming the Hamiltonian can destroy the string order
parameter (which is known to be fragile to small
perturbations\cite{Anfuso-2007}) and lift the degeneracy of the
edge states. Yet, as long as an appropriate set of symmetries is
preserved, the Haldane phase is stable\cite{Berg-2008,Gu-2009}, in
the sense that it is still separated from other, topologically
trivial phases by a thermodynamic phase transition in which the
gap closes. This stability, by itself, can be used as an
operational definition of the Haldane phase. However, it is
desirable to find a definition which can be stated in terms of the
ground state wavefunction of a single Hamiltonian. Recently, it
has been proposed that topological phases can be characterized by
their ``entanglement spectrum'', obtained by arbitrarily dividing
the system into two parts, tracing out one half and diagonalizing
the reduced density matrix of the
other.\cite{Haldane-2008,Li-2008,
Levin-2006,Kitaev-2006,Fidkowski-2009,Turner-2009b} This creates
artificial edges, without disrupting inversion symmetry.  For
example, the entanglement spectrum of the
Affleck-Kennedy-Lieb-Tasaki (AKLT) state\cite{Affleck-1987}
consists of two degenerate non-zero eigenvalues, which mimic the
doubly degenerate \emph{energy} edge spectrum of a system with a
physical boundary.\footnote{For critical system, studies of the entanglement spectrum has been shown to follow a universal distribution and does not reveal information about any non-local structure of the state.\cite{Calabrese-2008,Pollmann-2009c} In a recent study, the entanglement spectrum for particular, highly non-local cuts has been proposed to define a non-local order in gapless spin systems.\cite{Thomale-2009}}

In this paper, we show that the Haldane phase is characterized by a double
degeneracy of the entire entanglement spectrum. This degeneracy is caused by
the same set of symmetries which protect the stability of the Haldane phase,
applied to the eigenstates of the reduced density matrix.
If the Hamiltonian is deformed while keeping these symmetries intact, the
degeneracy remains until a phase boundary is crossed. This
symmetry-protected double degeneracy can be used to define the Haldane phase
in the most general situation, when both gapless edge states and a string
order parameter are absent.

The most surprising result of the analysis is that inversion
symmetry alone is enough to preserve the degeneracy of the
entanglement spectrum. If this is the only symmetry present, there
are no gapless edge modes, since edges break inversion symmetry.
There is also no string order, either.\footnote{It is possible to define a more general string order parameter that can be non-zero in this case.\cite{Cen09}}

This approach can be used to classify the phases of any one
dimensional system, given its symmetry group. For a given set of
symmetries, there are several types of gapped phases. One of them
is the \ \textquotedblleft non-degenerate\textquotedblright\ (or
``trivial'') phase, in which the eigenvalues of the density matrix
can be non-degenerate. Besides this, there can be several types of
``degenerate" (``non-trivial'') phases. The entanglement spectrum
in any one of the latter phases has at least two-fold degeneracy.
In this case, the density-matrix eigenstates transform in a
non-trivial way under a projective representation of the symmetry
group.

The entanglement spectrum, although being associated with a
partition of the system at a certain point in space, actually
carries highly non-local information about the ground state wave
function. We show that the double degeneracy of the entanglement
spectrum has a simple physical consequence. If one of the bonds of
the system is adiabatically weakened until its strength reaches
zero, the symmetry of the Hamiltonian continues to retain the
degeneracy in the entanglement spectrum across the weakened bond.
Hence, the von Neumann entropy of the partition in the final state
is equal to $\ln (2)$ once the bond is broken. This is a physical
reflection of the entanglement in the ground state, and can in
principle be used to identify it in experiment.

This paper is organized as follows. First we introduce in
Sec.~\ref{sec:mod} a spin-1 model Hamiltonian which has a Haldane
phase for a certain parameter range. In Sec.~\ref{sec:mps}, we
briefly review some properties of matrix-product states, which we
use to study properties of the entanglement spectrum, and show how
matrix-product states transform under symmetry operations. The
main result of this paper, namely the degeneracy of
the entanglement spectrum in the Haldane phase, is derived in Sec~\ref
{sec:DGDG}. Numerical results for several model Hamiltonians with
different symmetries are shown in Sec.~\ref{sec:exa}. In Sec.
\ref{sec:generalizations} we briefly outline the generalization of
these results towards a classification scheme of gapped phases in
one dimension. A more detailed discussion is deferred to a later
publication~\onlinecite{Turner-2009}. A numerical experiment,
which sheds light on the physical consequences of
the degeneracy of the entanglement spectrum, is presented in Sec.~\ref
{sec:evolve}. Finally, the results and conclusions are summarized in Sec.~
\ref{sec:sum}. Some details, concerning the application of the
above results to generalizations of the Haldane phase and a
derivation of the properties of the
Haldane phase under inversion, are discussed in the appendices.
\section{Model Hamiltonians\label{sec:mod}}

In order to study the stability of the Haldane phase, we will
mainly focus on various spin-1 model Hamiltonians with different
symmetries. As we will prove in Section \ref{sec:DGDG}, the
Haldane phase is protected by certain symmetries. One of them is a
symmetry under a bond-centered spatial inversion
\begin{equation}
S^{x,y,z}_{j}\rightarrow S^{x,y,z}_{-j+1},
\end{equation}

where $S^{x,y,z}_{j}$ are the spin-1 operators at site $j$. Other
possible symmetries are the time reversal (TR) symmetry
\begin{equation}
S_j^{x,y,z}\rightarrow-S_j^{x,y,z}
\end{equation}
or the symmetry with respect to spin rotations by $\pi$ about a pair of orthogonal axes. As long as \emph{at least one} of these
symmetries is not broken, the entire entanglement spectrum remains doubly
degenerate. Therefore, the Haldane phase maintains its identity and cannot
evolve adiabatically to another phase.

For concreteness, throughout most of this paper we consider the following
spin-1 model Hamiltonian
\begin{equation}
H_{0}=J \sum_{j} \vec{S}_{j} \cdot \vec{S}_{j+1} + B_x \sum_j S_j^x +U_{zz}
\sum_j (S^{z}_j)^2.  \label{H0}
\end{equation}
The symmetries of this model include translation, spatial inversion, a
rotation by $\pi$ around the $x$-axis, and a combination of a rotation by $%
\pi$ around the $y$-axis and time-reversal $e^{-i\pi S^y}\times$TR [which
takes $S_j^{x,z}\rightarrow S_j^{x,z},\ S_j^{y}\rightarrow-S_j^{y}$].

The phase diagram has been studied in Ref.~\onlinecite{Gu-2009}.
At large $U_{zz}$, we find a trivial insulator phase which can
described by a caricature state where all the sites are in the
$\left\vert S^{z}=0\right\rangle $ state. Furthermore, we find two
antiferromagnetic phases $Z_{2}^{y}$ and $Z_{2}^{z}$ with
spontaneous non-zero expectation values of $\langle S^{y}\rangle $
and $\langle S^{z}\rangle $, respectively. $U_{zz}=B_x=0$ is the
Heisenberg point, for which one finds the gapped Haldane phase.
Even when nonzero $U_{zz}$ and $B_x$ values are introduced into
this Hamiltonian, the Haldane phase is separated from the
\textquotedblleft non-degenerate phases\textquotedblright\ by
phase transitions. Here, we investigate the question of what
defines the Haldane phase and thus protects these transitions from
becoming smooth crossovers. We will later add different terms to
$H_0$ which break various symmetries, and show explicitly for
which classes of perturbations the Haldane phase remains
well-defined.

\section{Matrix product state representation\label{sec:mps}}

\subsection{Definitions}

In order to prove the above statements, we will use a matrix product state
(MPS) representation \cite{Fannes-1992} of the ground state wavefunction. We also use this
representation to compute the ground state properties numerically, using the
infinite time-evolving block decimation (iTEBD) method.\cite{Vidal-2007} The
iTEBD method is a descendent of the density matrix renormalization group
(DMRG) method.\cite{White-1992} For the sake of completeness, we now review
some of the properties of MPS's. A translationally invariant MPS for a chain
of length $L$ can formally be written in the following form\footnote{The discussion of the double degeneracy of the entanglement spectrum does not rely on translation symmetry. We impose translation symmetry only for simplicity.}
\begin{equation}
\left\vert \Psi \right\rangle =\sum_{\{m_{j}\}}\text{tr}\left[
\Gamma _{m_{1}}\Lambda \dots \Gamma _{m_{L}}\Lambda \right]
|m_{1}\dots m_{L}\rangle \text{.}  \label{mps}
\end{equation}
Here, $\Gamma _{m}$ are $\chi \times \chi $ matrices with $\chi $
being the the dimension of the matrices used in the MPS. The index
$m=-S,\dots ,S$ is the \textquotedblleft
physical\textquotedblright\ index, e.g., enumerating the spin
states on each site, and $\Lambda $ is a $\chi \times \chi $,
real, diagonal matrix. Ground states of one dimensional gapped
systems can be efficiently approximated by matrix-product
states\cite{Hastings-2007, Gottesman-2009, Schuch-2008}, in the sense that the
value of $\chi $ needed to approximate the ground state
wavefunction to a given accuracy converges to a finite value as $
N\rightarrow \infty$. We therefore think of $\chi$ as being a finite (but
arbitrarily large) number.

The matrices $\Gamma $, $\Lambda $ can be chosen such that they
satisfy the canonical conditions for an infinite MPS\cite{Vidal-2003a, Orus-2008}
\begin{equation}
\sum_{m}\Gamma^{\vphantom{\dagger }}_{m}\Lambda ^{2}\Gamma _{m}^{\dagger
}=\sum_{m}\Gamma _{m}^{\dagger }\Lambda ^{2}\Gamma^{\vphantom{\dagger }}
_{m}=\mathds{1}\text{.}  \label{canonical}
\end{equation}
These equations can be interpreted as stating that the transfer matrix
\begin{equation}
T_{\alpha \alpha ^{\prime };\beta \beta ^{\prime }}=\sum_{m}\Gamma _{m\beta
}^{\alpha }\left(\Gamma _{m\beta ^{\prime }}^{\alpha ^{\prime
}}\right)^{\ast}\Lambda _{\beta }\Lambda _{\beta ^{\prime }}
\label{transfer}
\end{equation}
has a right eigenvector $\delta _{\beta \beta ^{\prime }}$ with eigenvalue $
\lambda =1$. ($^\ast$~denotes complex conjugation.) Similarly, $
\tilde{T}_{\alpha \alpha ^{\prime };\beta \beta ^{\prime }}=\sum_{m}(\Gamma
_{m\beta ^{\prime }}^{\alpha ^{\prime }})^{\ast }\Gamma _{m\beta }^{\alpha
}\Lambda _{\alpha }\Lambda _{\alpha ^{\prime }}$ has a left eigenvector $
\delta _{\alpha \alpha ^{\prime }}$ with $\lambda =1$. We further
require that $\delta _{\alpha \alpha ^{\prime }}$ is the
\emph{only} eigenvector with eigenvalue $\left\vert \lambda
\right\vert \geq 1$ (which is equivalent to the requirement that
$|\psi\rangle$ is a pure state\cite{PerezGarcia-2008}).

The considerations given here become most intuitive when one considers,
formally, an infinite chain.

If the chain is infinite and has open ends, it may be partitioned at a
certain bond. The wavefunction can then be Schmidt decomposed \cite{Schmidt-1907} in the form
\begin{equation}
|\Psi \rangle =\sum_{\alpha }\lambda _{\alpha }|\alpha L\rangle |\alpha
R\rangle ,  \label{schmidt}
\end{equation}
where $|\alpha L\rangle $ and $|\alpha R\rangle $ ($\alpha =1,\dots ,\chi $)
are orthonormal basis vectors of the left and right partition, respectively.
In the limit $L\rightarrow \infty $, and under the canonical conditions (\ref
{canonical}), the Schmidt eigenvalues $\lambda _{\alpha }$ are
simply the entries of the $\Lambda $ matrix, $\Lambda _{\alpha
\alpha}$. $\lambda _{\alpha }^{2}$ are the eigenvalues of the
reduced density matrix of either of the two partitions, and are
referred to as the \emph{entanglement spectrum}. The entanglement
entropy is $S=-\sum_{\alpha }\lambda _{\alpha }^{2}\ln \lambda
_{\alpha }^{2}$. This corresponds to the von Neumann entropy of
the reduced density matrix. The states $|\alpha L\rangle$ and
$|\alpha R\rangle$ can be obtained by multiplying together all the
matrices to the left and right of the bond, e.g., if the broken
bond is between sites $0$ and $1$, $|\alpha
L\rangle=\sum_{\{m_{j}\},j\leq0}\left[\prod_{k\leq 0} \Lambda
\Gamma
_{m_{k}} \right]_{\gamma\alpha} |\dots m_{-2}m_{-1}m_0\rangle$. Here, $
\gamma $ is the index of the row of the matrix; when the chain is infinitely
long, the value of $\gamma$ affects only an overall factor in the
wavefunction. Reviews of MPS's as well as the canonical form can be found in
Refs.~\onlinecite{PerezGarcia-2007,Orus-2008}.

\subsection{Symmetries in matrix product states}

In order to study the consequences of symmetries of the wavefunctions, it is
useful to first study how these symmetries are reflected in the MPS\
representation. If $|\Psi \rangle $ is invariant under a \emph{local symmetry
} which is represented in the spin basis as a unitary matrix $\Sigma
_{mm^{\prime }}$, then the $\Gamma $ matrices can be shown to satisfy\cite
{PerezGarcia-2008}

\begin{equation}
\sum_{m^{\prime }}\Sigma _{mm^{\prime }}\Gamma _{m^{\prime }}=e^{i\theta
_{\Sigma }}U_{\Sigma }^{\dagger }\Gamma _{m}U^{\vphantom{\dagger }} _{\Sigma
}\text{,}  \label{trans}
\end{equation}
where $U_{\Sigma }$ is a unitary matrix which commutes with the $\Lambda $
matrices, and $e^{i\theta _{\Sigma }}$ is a phase. Thus, the matrices $
U_{\Sigma }$ form a $\chi-$dimensional (projective) representation
of the symmetry group of the wavefunction. In close analogy to the
derivation in Ref.~\onlinecite {PerezGarcia-2008}, we can derive a
similar relation to Eq. (\ref{trans}) for time reversal and
inversion symmetry. For a time reversal transformation $\Gamma
_{m^{\prime }}$ is replaced by $\Gamma _{m^{\prime }}^{\ast }$
(complex conjugate) on the left hand side. Finally, in the case of
inversion symmetry $\Gamma _{m^{\prime }}$ is replaced by $\Gamma
_{m^{\prime }}^{T}$ (transpose) on the left hand side of Eq.
(\ref{trans}).

\section{Degeneracies in the entanglement spectrum\label{sec:DGDG}}

We now turn to derive our main result, namely the degeneracies in the
entanglement spectrum of the wavefunction in the Haldane phase. Our strategy
is to determine when the transformation law for the Schmidt eigenstates
under the symmetry operations of the system is non-trivial. From Eq. (\ref
{trans}), the Schmidt eigenstates of the left half of the system, $\vert
\alpha L \rangle$, transform under a symmetry operation $\Sigma$ according
to the following rule:

\begin{equation}
\Sigma|\alpha L\rangle=\sum_\beta(U_\Sigma)_{\beta\alpha}|\beta L\rangle.
\label{eq:starsontrees}
\end{equation}
Similarly, the right Schmidt states $\vert \alpha R \rangle$
transform by the conjugate matrix. Thus, the Schmidt eigenstates
transform according to a projective representation of the symmetry
group of the system. The phases of the matrices $U_\Sigma$ are not
uniquely determined by Eq. (\ref{trans}), or by Eq.
(\ref{eq:starsontrees}).
The phase ambiguities turn out to be the
key to proving the degeneracies of the entanglement spectrum. We
will show that for certain symmetries, there can be situations
where \emph{the irreducible representations present in $U_\Sigma$
are all multi-dimensional}. In these cases, which are identified
with the Haldane phase (or a generalization of it), the entire
entanglement spectrum has non-trivial degeneracies.

\subsection{Inversion symmetry}

\label{inv} As a first example, let us consider a system which is symmetric
under spatial inversion. The transformation law of $\Gamma $ is written as

\begin{equation}
\Gamma _{m}^{T}=e^{i\theta _{\mathcal{I}}}U_{\mathcal{I}}^{\dagger }\Gamma
_{m}U^{\vphantom{\dagger }}_{\mathcal{I}}\text{,}  \label{reflect}
\end{equation}
where $U_{\mathcal{I}}$ is a unitary matrix and $\theta _{\mathcal{I}}\in
\lbrack 0,2\pi )$ is a phase. Iterating this relation twice gives

\begin{equation}
\Gamma _{m}=e^{2i\theta _{\mathcal{I}}}\left( U^{\vphantom{\ast}}_{\mathcal{I
}}U_{\mathcal{I}}^{\ast }\right) ^{\dagger }\Gamma _{m}U^{\vphantom{\ast }}_{
\mathcal{I}}U_{\mathcal{I}}^{\ast }\text{.}
\end{equation}
Now, the relation implies that
\begin{equation}
\sum_{m}\Gamma _{m}^{\dagger }\Lambda U^{\vphantom{\ast }}_{\mathcal{I}}U_{
\mathcal{I}}^{\ast }\Lambda \Gamma _{m}=e^{2i\theta _{\mathcal{I}}}U_{
\mathcal{I}}^{\vphantom{\ast}}U_{\mathcal{I}}^{\ast }\text{,}  \label{double}
\end{equation}%
where we have used Eq. (\ref{canonical}) and the fact that $[U_{\mathcal{I}%
},\Lambda]=0$. Thus $U^{\vphantom{\ast }}_{\mathcal{I}}U_{\mathcal{I}}^{\ast
}$ is an eigenvector of the transfer matrix $T$ [Eq. (\ref{transfer})] with
eigenvalue $e^{2i\theta _{\mathcal{I}}}$. Since by our assumption, the only
unimodular eigenvalue of $T$ is $\lambda =1$ and this eigenvalue is unique,
we find that $e^{2i\theta _{\mathcal{I}}}=1$ and $U^{\vphantom{\ast }}_{%
\mathcal{I}}U_{\mathcal{I}}^{\ast }=e^{i\phi _{\mathcal{I}}}\mathds{1}$
where $\phi _{\mathcal{I}}$ is a phase. Hence $U_{\mathcal{I}}^{T}=U_{%
\mathcal{I}}e^{-i\phi _{\mathcal{I}}}$. Repeating this relation twice, we
find that $e^{-2i\phi _{\mathcal{I}}}=1$, i.e. $\phi _{\mathcal{I}}=0$ or $%
\pi $.

If $\phi _{\mathcal{I}}=\pi $, then $U_{\mathcal{I}}$ is an antisymmetric
matrix. From this we find that all the eigenvalues $\Lambda _{\alpha}$ are
\emph{at least} doubly degenerate. Moreover, the corresponding multiplicity $%
k_{\alpha}$ is even for all $\alpha$. This follows from the fact that $U_{%
\mathcal{I}}$ transforms the $k_{\alpha}$-dimensional subspace of states
with eigenvalue $\Lambda_{\alpha}$ within itself. Therefore, the matrix $U_{%
\mathcal{I}}^{\alpha}$ (projected into subspace $\alpha$) satisfies $\det
U^{\alpha}_{\mathcal{I}}=\det [(U^{\alpha}_{\mathcal{I}})^{T}]=\det \left(
-U^{\alpha}_{\mathcal{I}}\right) =\left( -1\right) ^{k_{\alpha}}\det
U^{\alpha}_{\mathcal{I}}$. But since $U^{\alpha}_{\mathcal{I}}$ is unitary, $%
\det U^{\alpha}_{\mathcal{I} }\neq 0$ and therefore $\left( -1\right)
^{k_{\alpha}}=1$.

The fact that, in the presence of inversion symmetry, the phase $\phi_{%
\mathcal{I}}$ can only take discrete values ($0$ or $\pi $), leads to phase
transitions between states when one would not expect them on the basis of
the Landau paradigm of broken symmetry. If an inversion-symmetric
wavefunction evolves continuously, its characteristic phase $\phi _{\mathcal{%
I}}$ cannot change discontinuously, and therefore its value is fixed. The
only way $\phi _{\mathcal{I}}$ can change is through a critical point, where
either the correlation length diverges because the transfer matrix $T$ has a
pair of unimodular eigenvectors (and the main relation, $U_I^*U_I^{%
\vphantom{\ast }}=e^{i\phi_I}\mathds{1}$ cannot be proven) or
there is simply a discontinuous change in the ground state
wavefunction (\emph{i.e.}, a first order transition). We can
therefore identify two distinct states,
characterized by $\phi _{\mathcal{I}}=0,\pi $. The state with $\phi _{%
\mathcal{I}}=\pi $ can be identified with the Haldane phase. To show this,
we consider the AKLT\ state with $\Gamma _{a}=\sigma _{a}$, $\Lambda =\frac{1%
}{\sqrt{2}}\mathds{1}$, where $\sigma _{a}$ ($a=x,y,z $) are Pauli matrices,
and we use the time-reversal invariant spin basis $\left\vert x\right\rangle
=\frac{1}{\sqrt{2}}\left( \left\vert 1\right\rangle -\left\vert
-1\right\rangle \right) $, $\left\vert y\right\rangle =\frac{i}{\sqrt{2}}%
\left( \left\vert 1\right\rangle +\left\vert -1\right\rangle \right) $, $%
\left\vert z\right\rangle =\left\vert 0\right\rangle $. Under inversion, $%
\sigma _{a}\rightarrow \sigma _{a}^{T}=-\sigma _{y}\sigma _{a}\sigma _{y}$
and thus $U_{\mathcal{I}}=\sigma _{y}$ and $\theta_\mathcal{I}=\pi$. Since $%
\sigma^{\vphantom{\ast }} _{y}\sigma _{y}^{\ast }=-\mathds{1}$, we find that $%
e^{i\phi _{\mathcal{I}}}=-1$. The AKLT\ state is known to describe the same
phase as the Haldane phase.\cite{Affleck-1987} Therefore, we conclude that
the Haldane phase is characterized by $e^{i\theta _{\mathcal{I}}}=-1$, $\
e^{i\phi _{\mathcal{I}}}=-1$, and a doubly degenerate entanglement spectrum.
The wave function cannot evolve continuously if the phases $\theta _{%
\mathcal{I}}$ or $\phi _{\mathcal{I}}$ change discontinuously. This implies
that changes of $\theta _{\mathcal{I}}$ or $\phi _{\mathcal{I}}$ between $0$
and $\pi$ are always accompanied by a phase transition. Consequently, the
degeneracy in the Haldane phase can only be lifted by a phase transition.
The full argument for the existence of a transition in such case appears in
Ref.~\onlinecite{Pollmann-2009a}.

In the discussion above, we have assumed that the system is invariant under
both inversion and translation [see Eq. (\ref{mps})]. However, in fact,
inversion symmetry alone is sufficient to protect the double degeneracy in
the entanglement spectrum, as long as it is bond-centered. To show this, one
can imagine adding a general commensurate perturbation to the Hamiltonian,
such that the unit cell is enlarged. One can still write the ground state
wave function in a translationally invariant form (\ref{mps}) were each site
represents a single unit cell. If the unit cell is defined such that it ends
at an inversion-symmetric bond, the new system is also inversion symmetric,
and the entanglement spectrum degeneracy remains protected. Since the size
of the unit cell can be arbitrarily large, it is clear that translational
symmetry cannot be essential for this argument to hold. The same argument
can be made in the case of local symmetries, such as the ones described in
Sections \ref{TRS},\ref{rot}.

\subsection{Time reversal symmetry}

\label{TRS}

The transformation of the MPS\ wavefunction $\Gamma $ matrices under TR has
the form
\begin{equation}
\sum_{m^{\prime }}\left( \Sigma _{\mathcal{T}}\right) _{mm^{\prime }}\Gamma
_{m^{\prime }}^{\ast }=e^{i\theta _{\mathcal{T}}}U_{\mathcal{T}}^{\dagger
}\Gamma _{m}U^{\vphantom{\dagger }}_{\mathcal{T}}\text{.}  \label{SR}
\end{equation}
Here we have used the $S^z$ basis for the spins ($m=-1,0,1$), and $\Sigma_{%
\mathcal{T}}=e^{i\pi S^{y}}$. From this, one can derive (in close analogy
with the case of spatial inversion) that $U^{\vphantom{\ast }}_{\mathcal{T}%
}U_{\mathcal{T}}^{\ast }=e^{i\phi _{\mathcal{T}}}\mathds{1}$ where $\phi _{%
\mathcal{T}}$ can be either $0$ or $\pi $. If $e^{i\phi _{\mathcal{T}}}=-1$,
then the double degeneracy of the entanglement spectrum follows (precisely
as in the previous section). The AKLT\ state $\Gamma $ matrices transform as
$\Gamma_{m}\rightarrow \sigma_{y}\Gamma_{m}\sigma _{y}$. Thus $U_{\mathcal{T}%
}=\sigma_y$ and time reversal symmetry is sufficient to protect the double
degeneracy of the entanglement spectrum in the Haldane phase.

The transformation of the MPS\ wavefunction $\Gamma $ matrices under $%
e^{-i\pi S^{y}}\times$TR corresponds to a complex conjugation ($\mathcal{C}%
\mathcal{C}$) of the wavefunction and has the form
\begin{equation}
\Gamma_m^{\ast }=e^{i\theta _{\mathcal{C}\mathcal{C}}}U_{\mathcal{C}\mathcal{%
C}}^{\dagger }\Gamma _{m}U^{\vphantom{\dagger }}_{\mathcal{C}\mathcal{C}}%
\text{.}  \label{TR}
\end{equation}%
From this, one can derive that $U_{\mathcal{C}\mathcal{C}}U_{\mathcal{C}%
\mathcal{C}}^{\ast }=e^{i\phi _{\mathcal{C}\mathcal{C}}}\mathds{1}$ where $%
\phi _{\mathcal{C}\mathcal{C}}$ can be either $0$ or $\pi $. The AKLT\ state
$\Gamma $ matrices transform as $\Gamma_{m}\rightarrow \Gamma_{m}$. It
follows that $U_{\mathcal{C}\mathcal{C}}=\mathds{1}$ and thus $e^{-i\pi
S^{y}}\times$TR alone \emph{is not sufficient} to protect the double
degeneracy of the entanglement spectrum in the Haldane phase. Physically,
this means that it is possible to add to the Hamiltonian a perturbation
which is invariant under $e^{-i\pi S^{y}}\times$TR but destroys the Haldane
phase, in the sense that it is no longer separated from an unentangled
product state by a phase transition. This perturbation has to break all
other symmetries that may protect the Haldane phase (an example of such a
perturbation can be found in Sec.~\ref{sec:exa}).

\subsection{Sets of Rotations}

\label{rot}

A symmetry of rotation about a single axis by $\frac{2\pi}{n}$, where $n$ is
an integer, does not lead to any classification of phases. If $\Sigma$ is a
rotational symmetry of order $n$, one can show that $U_{\Sigma}^n=e^{i\phi}$
as in the previous section. Rescaling $U_\Sigma$ by $e^{\frac{i\phi}{n}}$
leaves Eq. (\ref{trans}) satisfied, and shows that $\phi $ has no
significance. However, when there are multiple symmetries, there is also a
phase for each pair of symmetries $\Sigma_1,\Sigma_2$. This phase is defined
by noting that the transformation of Schmidt states corresponding to $%
\Sigma_1\Sigma_2$ can differ by a phase from the product of the Schmidt
state representations of $\Sigma_1$ and $\Sigma_2$:
\begin{equation}
U_{\Sigma_1}U_{\Sigma_2}=e^{i\rho(\Sigma_1,\Sigma_2)}U_{\Sigma_1\Sigma_2}.
\label{eq:schur}
\end{equation}
If the phase $\rho(\Sigma_1,\Sigma_2)$ cannot be gauged away by redefining
the phases of $U_{\Sigma_{1,2}}$, then the combined symmetry can lead to a
protected Haldane phase.

A concrete example is a system with symmetry under the dihedral
group D$_2$ of $\pi$ rotations about three orthogonal axes (say,
$x,y$ and $z$ axes). 
Since the product of $\pi$
rotations about the $x$ and $z$ axes $\mathcal{R}_x \mathcal{R}_z$
or $\mathcal{R}_z \mathcal{R}_x$ give a $\pi$ rotation about the
$y$ axis $\mathcal{R}_y$, the group is equivalent to $Z_2 \times
Z_2$.
 Thus it is sufficient to consider the action of two
generators, say $\mathcal{R}_{x}$ and $\mathcal{R}_{z}$. For
$\mathcal{R}_{x}$,
\begin{equation}
\sum_{m^{\prime }}\left( \Sigma _{x}\right) _{mm^{\prime }}\Gamma
_{m^{\prime }}=e^{i\theta _{x}}U_{x}^{\dagger }\Gamma _{m}U^{%
\vphantom{\dagger}}_{x}\text{,}
\end{equation}%
where $\Sigma _{x}=e^{i\pi S^{x}}$. Repeating this relation twice, we get $%
\Gamma _{m}=e^{2i\theta _{x}}\left( U_{x}^{\dagger }\right) ^{2}\Gamma
_{m}U_{x}^{2}$. From this it follows [analogously to the arguments below Eq.
(\ref{double})] that $e^{2i\theta _{x}}=1$ and $U_{x}^{2}=e^{i\phi _{x}}%
\mathds{1}$. The phase factor $e^{i\phi _{x}}$ is not important, since it
can be absorbed in $U_{x}$. Therefore we can assume that $U_{x}^{2}=%
\mathds{1}$. Similarly for $\mathcal{R}_{z}$, we arrive at $U_{z}^{2}=%
\mathds{1}$. The combined operation $\mathcal{R}_{x}\mathcal{R}_{z}$,
however, may give rise to a non-trivial phase factor. By repeating this
symmetry twice, the associated unitary matrix $U_{x}U_{z}$ can be shown (in
the same way as above) to satisfy $U_{x}U_{z}=e^{i\phi _{xz}}U_{z}U_{x}$.
Since the phases of $U_{x}$ and $U_{z}$ have been defined, the phase factor $%
e^{i\phi _{xz}}$ is \emph{not} arbitrary, and can have a physical meaning.
Clearly $e^{i\phi _{xz}}=\pm 1$. If $e^{i\phi _{xz}}=-1$, then the spectrum
of $\Lambda $ is doubly degenerate, since $\Lambda $ commutes with the two
unitary matrices $U_{x}$, $U_{z}$ which anti-commute among themselves. For
the AKLT state, $U_{x}=\sigma _{x}$ and $U_{z}=\sigma _{z}$, therefore $%
U_{x}U_{z}=-U_{z}U_{x}$, and the Haldane phase is protected if the system is
symmetric under both $\mathcal{R}_{x}$ and $\mathcal{R}_{z}$.

\section{Examples\label{sec:exa}}

We now demonstrate how the symmetries discussed above stabilize
the Haldane phase. We use the iTEBD method to numerically
calculate the ground state of the model given by Eq. (\ref{H0}),
augmented by various symmetry-breaking perturbations. We used
MPS's with a dimension of $\chi=80$ for the simulations. The
double degeneracy of the entanglement spectrum is used to identify
the Haldane phase.

\begin{figure}[htp!]
\begin{center}
\includegraphics[width=85mm]{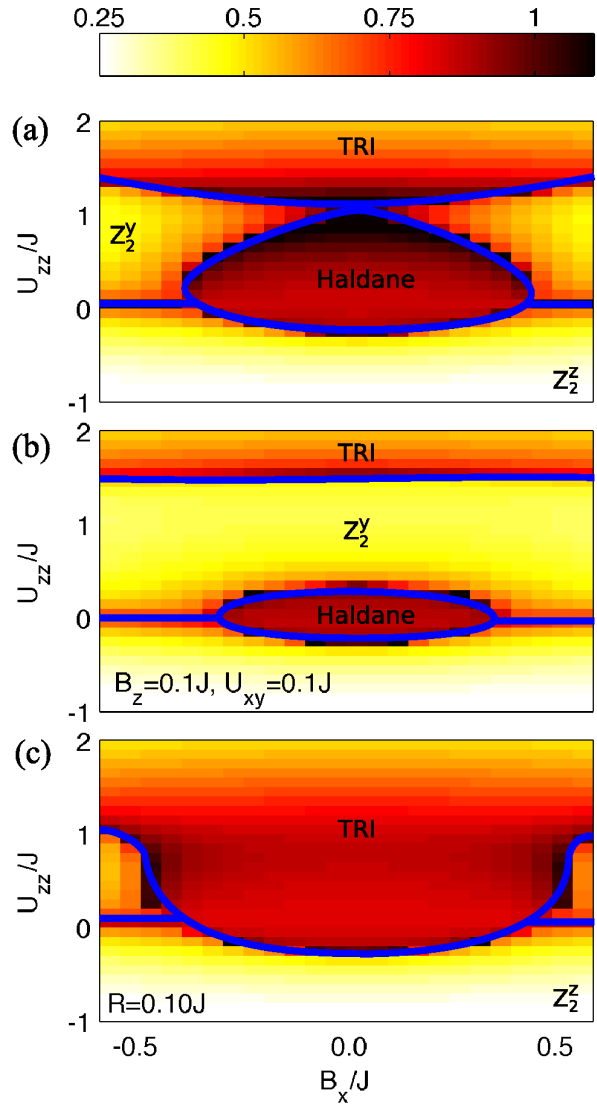}
\end{center}
\caption{The colormaps show the entanglement entropy $S$ for different
spin-1 models: Panel (a) shows the data for Hamiltonian $H_0$ in (\protect
\ref{H0}), panel (b) for $H_0$ plus a term which breaks the $e^{-i\protect%
\pi S^{y}}\times$TR symmetry [Eq. (\protect\ref{eq:break_TR})], and panel
(c) for $H_0$ plus a term which breaks the $e^{-i\protect\pi S^{y}}\times$TR
and inversion symmetry [Eq. (\protect\ref{eq:break_inv})]. The blue lines
indicate a diverging entanglement entropy as a signature of a continuous
phase transition. The phase diagrams contain four different phases: A
trivial insulating phase (TRI) for large $U_{zz}$, two symmetry breaking
antiferromagnetic phases $Z_2^z$ and $Z_2^y$, and a Haldane phase (which is
absent in the last panel).}
\label{fig:pd_S}
\end{figure}

\begin{figure}[htp!]
\begin{center}
\includegraphics[width=85mm]{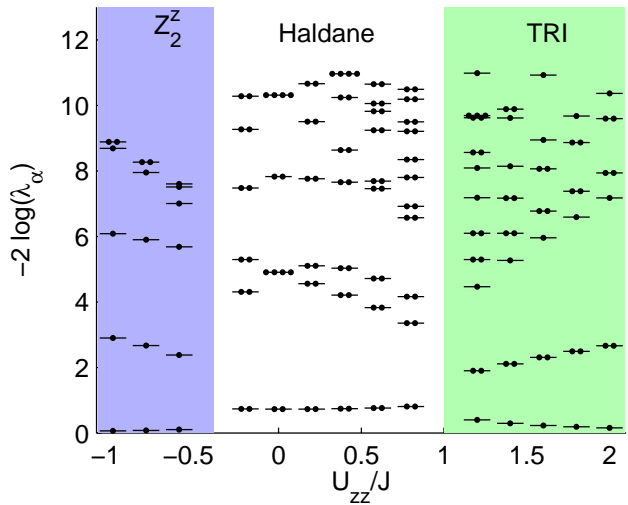}
\end{center}
\caption{Entanglement spectrum of Hamiltonian $H_0$ in (\protect\ref{H0})
for $B_x=0$ (only the lower part of the spectrum is shown). The dots show
the multiplicity of the Schmidt values, which is even in the entire Haldane
phase.}
\label{fig:es}
\end{figure}

\begin{figure}[tbp]
\begin{center}
\includegraphics[width=85mm]{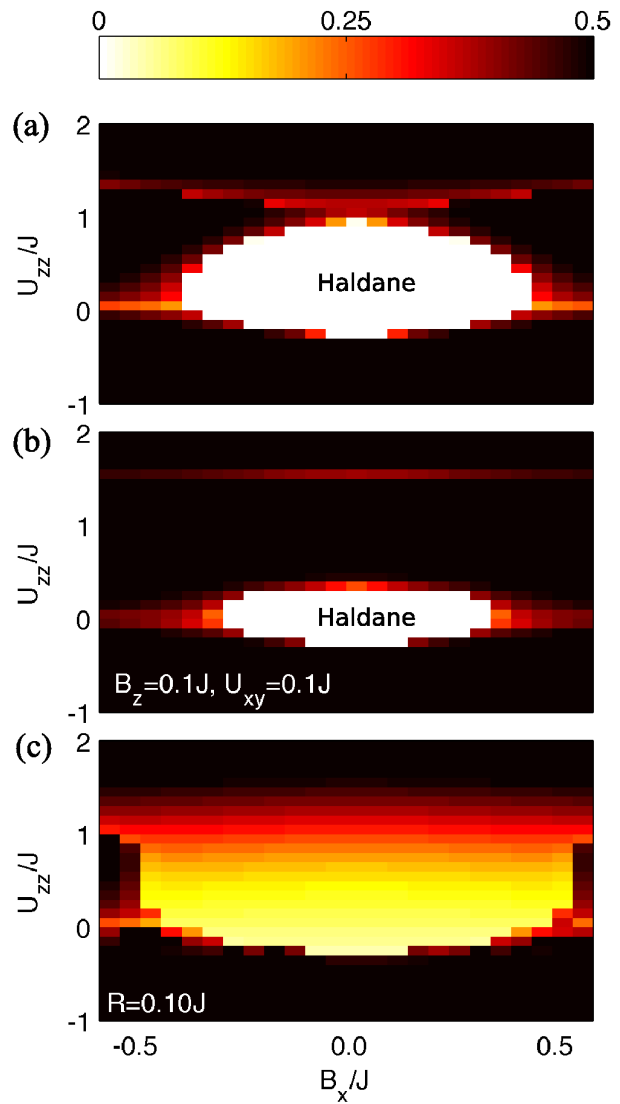}
\end{center}
\caption{The colormaps show the difference between the two largest Schmidt
values $|\protect\lambda_1-\protect\lambda_2|$ for different spin-1 models.
Panel (a) corresponds to the original Hamiltonian $H_0$ in (\protect\ref{H0}%
), panel (b) to $H_0$ plus a term that breaks the time reversal symmetry
[Eq. (\protect\ref{eq:break_TR})], and panel (c) to $H_0$ plus a term which
breaks time reversal and inversion symmetry [Eq. (\protect\ref{eq:break_inv}%
)]. The quantity $|\protect\lambda_1-\protect\lambda_2|$ is zero only in the
Haldane phase.}
\label{fig:pd_gap}
\end{figure}

\paragraph*{Example 1:}

We begin with the original Hamiltonian $H_0$ in (\ref{H0}). This
model is translation invariant, invariant under spatial inversion,
under $e^{-i\pi S^{x}}$, and under $e^{-i\pi S^{y}}\times$TR.
Using the above argument, we know that inversion symmetry alone is
sufficient to protect a Haldane phase. The phase diagram is shown
in FIG.~{\ref{fig:pd_S}}(a) and agrees with the results of
Ref.~\onlinecite{Gu-2009}. A diverging entanglement entropy
indicates a phase transition (see for example
Ref.~\onlinecite{Vidal-2003}) and we use observables such as
$S_y$, $S_z$ to reveal the nature of the phases. In our approach,
the Haldane phase can be easily identified by looking at the
degeneracy of the entanglement spectrum as shown in
FIG.~{\ref{fig:es}}: the even degeneracy of the entanglement
spectrum occurs in the entire phase.

In the entanglement spectrum of the TRI phase, there are both
singly and multiply-degenerate levels. We have checked that the
multiple degeneracies in the TRI spectrum can be lifted by adding
symmetry-breaking perturbations to the Hamiltonian (while
preserving inversion symmetry). In the Haldane phase, on the other
hand, the double degeneracy of the entire spectrum is robust to
adding such perturbations.

The colormap in FIG.~{\ref{fig:pd_gap}}%
(a) shows the difference of the two largest Schmidt values for the whole $%
B_x $-$U_{zz}$ phase diagram. In the Haldane phase, the whole spectrum is at
least two-fold degenerate and thus the difference is zero.

\paragraph*{Example 2:}

The Hamiltonian $H_0$ has in fact more symmetries than are needed to
stabilize the Haldane phase. To demonstrate this, we add a perturbation $H_1$
of the form
\begin{equation}
H_1=B_z\sum_jS^z_j+U_{xy} \sum_j\left( S_j^x S_j^y+ S_j^y S_j^x \right).
\label{eq:break_TR}
\end{equation}
$H_1$ is translation invariant and symmetric under spatial inversion, but
breaks the $e^{-i\pi S^{x}}$ and the $e^{i\pi S^y}\times$TR symmetry. The
phase diagram for fixed $B_z=0.1J$ and $U_{xy}=0.1J$ as a function of $B_x$
and $U_{zz}$ is shown in FIG.~{\ref{fig:pd_S}}(b). As predicted by the
symmetry arguments above, we find a finite region of stability for the
Haldane phase. This region is characterized, as before, by a twofold
degeneracy in the entanglement spectrum, as shown in FIG.~{\ref{fig:pd_gap}}%
(b).

\paragraph*{Example 3:}

In this example, we consider a case in which there is no symmetry that
protects the Haldane phase. We add the following inversion symmetry-breaking
term:
\begin{eqnarray}
H_{1} &=&R\sum_{j} [ S_{j}^{z}(S_{j}^{x}S_{i+1}^{x}+S_{j}^{y}S_{j+1}^{y})
\nonumber \\
&&-S_{j+1}^{z}(S_{j}^{x}S_{j+1}^{x}+S_{j}^{y}S_{j+1}^{y})+\text{H.c.}]\text{.%
}  \label{eq:break_inv}
\end{eqnarray}%
Note that this term is invariant under $e^{i\pi S^y}\times$TR. The phase
diagram for the parameter $R=0.1J$ is shown in FIG.~{\ref{fig:pd_S}}(c). As
predicted by the symmetry arguments above, we do \emph{not} find a Haldane
phase with a twofold degeneracy (see FIG.~{\ref{fig:pd_gap}}(c)). The
Haldane phase region is continuously connected to the TRI phase. The same
scenario appears if we consider very small $R$.

\paragraph*{Example 4:}

Another example in which the Haldane phase and the TRI phase are
continuously connected is recently given in Eqn.~(6) of
Ref.~\onlinecite{Nakamura-2009}. Their model also does not have
any of the symmetries which protects the Haldane phase. Thus their
finding is consistent with our analysis.

\section{Classification of gapped phases in one dimension}

\label{sec:generalizations} In Sec. \ref{sec:DGDG} we have identified
several $\phi$--parameters, such as $\phi_{\mathcal{I}}$, $\phi_{\mathcal{T}%
} $, and $\phi_{xz}$, which parametrize the phase ambiguities in the
symmetry operations acting on the Schmidt eigenstates of the wavefunction.
When one of these parameters is nonzero, the entanglement spectrum is
degenerate, and a non-trivial (Haldane-like) ``degenerate'' phase is stable
over a finite range in parameter space.

When more than one symmetry is present, the non-trivial (``degenerate'')
phases may be classified into several families depending on the combination
of values taken by the corresponding $\phi$'s. In fact, there are even more
phases than this argument would naively suggest.\cite{Turner-2009} Most
generally, given the symmetry group of the system, the phases can be
classified according to all the possible in-equivalent projective
representations of the symmetry group.
The general classification scheme of one
dimensional gapped phases will not be described in detail in this work, but
will be deferred to a later publication Ref.~\onlinecite{Turner-2009}.

In Appendix.~\ref{sec:gen}, we consider various generalizations of
the Haldane phase, which are protected by different symmetries. In
a $S=1$ antiferromagnetic chain with Dzyaloshinskii-Moriya
interactions in a magnetic field, we show that there is a stable
Haldane-like phase which is protected by a modified inversion
symmetry. The extended Bose-Hubbard model also has a generalized
Haldane phase.\cite{DallaTorre-2006} This phase is shown to be
inequivalent to the usual Haldane phase of spin-1 chains, which is
also supported by the symmetry group of this system.

\section{Adiabatic bond weakening\label{sec:evolve}}

\begin{figure}[tbp]
\begin{center}
\includegraphics[width=85mm]{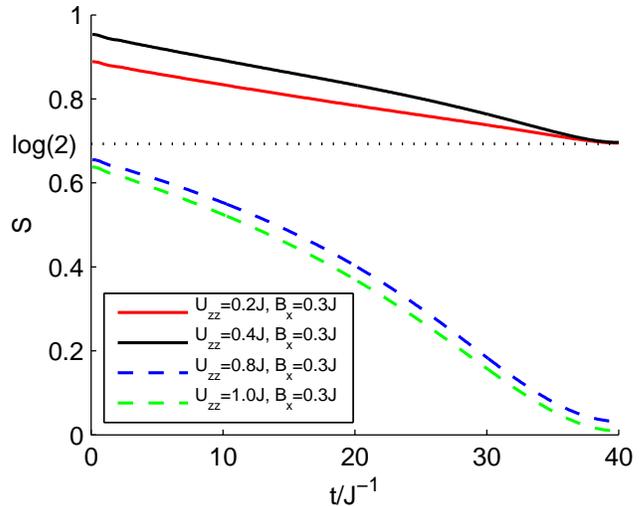}
\end{center}
\caption{Half-chain entanglement entropy of the model Hamiltonian (\protect\ref{H0}) at
a bond which is slowly weakened as a function of time $J_{\text{weak}%
}=J-t\Gamma$ and $\Gamma=J^2/40$. The entanglement entropy of the resulting
state is $\ln2$ in the Haldane phase and zero otherwise.}
\label{fig:weak}
\end{figure}

The doubly degenerate entanglement spectrum is a unique feature of the
Haldane phase, which can be used to distinguish it from other phases.
However, since the entanglement spectrum is a highly non-local property, one
may wonder whether it has any physical consequences, which can be accessed
in experiments.

In the introduction, we discussed an adiabatic process in which a single
bond in the system is slowly weakened to zero. The degree of correlation
remaining across this bond can be measured by the entanglement entropy. We
will show that, if the system starts in the Haldane phase and inversion
symmetry about the weakened bond is preserved throughout the process, the
minimum value of the entanglement entropy of the two halves is $\ln(2)$.
This is because the entanglement spectrum eigenvalues remain doubly
degenerate. The minimum entropy is reached if just one pair of entanglement
eigenvalues is nonzero.

{This property of the Haldane phase can be used, in principle, to
identify it in an experiment. It means that, starting from the
Haldane phase and separating it adiabatically into two halves $A$
and $B$, some degree of entanglement between $A$ and $B$ must
remain in the final state, as long as the symmetry that protects
the Haldane phase is respected. This manifests itself in physical
correlation functions between the two halves. Namely, there must
exist a pair of physical operators $\mathcal{O}_A$ and
$\mathcal{O}_B$  belonging to subsystems A and B, respectively,
such that the disconnected correlation function $C_{A,B}=\langle
\mathcal{O}_A \mathcal{O}_B \rangle - \langle \mathcal{O}_A
\rangle \langle \mathcal{O}_B \rangle$ remains non-zero, even when
the two subsystems are completely disconnected. Starting from a
``non-degenerate'' phase, on the other hand, the final state can
be completely unentangled across the cut bond.\footnote{In
particular cases, it is possible that the final state has some
residual entanglement even if the initial state is Ònon-
degenerateÓ, e.g., due to an accidental level crossing in the
bond-weakening process. However, this situation is non-generic,
and therefore probably unstable to small perturbations.}}


To simulate the weakening of one bond numerically, we start by
preparing ground states of Hamiltonian (\ref{H0}) for different
parameters using the iTEBD algorithm with a unit cell of size
$L=80$ which is large compared to the correlation length.
{We then evolve this state in time while
decreasing the strength of one bond in between two half chains,
$J_{\text{weak}}$, according to $ J_{\text{weak}}=J-t\Gamma$.} For
the rate $\Gamma=J^2/40$ and $ t=0\dots40J^{-1}$, we found that
this time evolution is essentially adiabatic. We calculate the
entanglement entropy at the middle bond as a function of time. The
result for $B_x=0.3J$ and various values for $U_{zz}$ is shown in
FIG~\ref{fig:weak}. Within the Haldane phase ($U_{zz}=0.2J$ and $
U_{zz}=0.4J$) the entanglement entropy at the end of the weakening
process is equal to $\ln(2)$. In the TRI phase ($U_{zz}=0.8J$ and
$U_{zz}=1.0J$), the entanglement entropy decreases monotonically
to zero.

The robustness of the degeneracy in the entanglement spectrum of
the Haldane phase has an intuitive explanation, as follows. Let us examine the Schmidt decomposition of the ground state
wavefunction corresponding to dividing the system along the weakened bond.
Initially, since $\phi _{\mathcal{I}}=\pi$, the Schmidt states
appear in doublets. As we show in Appendix B, the Schmidt
decomposition can be written as
\begin{equation}
|\Psi \rangle =\sum_{\alpha=1 }^{\chi/2}\lambda _{2\alpha-1}(|\alpha ,1\rangle |\overline{
\alpha ,2}\rangle -|\alpha ,2\rangle |\overline{\alpha ,1}\rangle
)\text{,} \label{Sch_asym}
\end{equation}
where $\lambda _{2\alpha-1}$ are the Schmidt eigenvalues, $|\alpha ,i\rangle $
with $i=1,2$ are Schmidt states of the left subsystem, and $\overline{
|\alpha ,i\rangle }=\mathcal{I}|\alpha ,i\rangle $ are
the inversion-related states on the right subsystem.
$|\Psi \rangle $ is odd under inversion, as can be seen by
applying the inversion operator $\mathcal{I}$. Since the
Hamiltonian remains symmetric under inversion throughout the bond
weakening process, $|\Psi \rangle $ has to remain antisymmetric.
Thus, at the end of the adiabatic evolution, the system is in the
ground state \emph{of the antisymmetric sector}, which generally
differs from the true ground state (unless the ground state of
each of the disconnected halves is degenerate). This is because
the true ground state is symmetric under inversion. The
antisymmetric sector ground state can be written as
$\frac{1}{\sqrt{2}}(|0\rangle \overline{|1\rangle} -|1\rangle
\overline{|0\rangle})$. Here $|0\rangle $ and $|1\rangle $ are,
respectively, the ground state and
first excited state of the left subsystem, and $\overline{|0\rangle }$,$
\overline{|1\rangle }$ are the ground state and first excited state of
the right subsystem, which are related by inversion to the
corresponding states on the left. Thus the entanglement spectrum remains doubly degenerate, and
entanglement entropy in the final state is $\ln \left( 2\right) $.

Note that this property of the Haldane phase is not associated
with the existence of zero energy edge states. (The two states
$|0\rangle $ and $|1\rangle $ do not have to be degenerate.) In
particular, the Hamiltonian $ H_{0}$ [Eq. (\ref{H0})] does not
have any zero-energy edge state at an open boundary.

\section{Summary\label{sec:sum}}

In this work, we have considered the Haldane phase of $S=1$ chains as an
example of a ``topological'' phase in one dimension. It has been known for a
long time that this phase cannot be characterized by any local
symmetry-breaking order parameter, and that its unusual character only shows
up in non-local properties, such as zero-energy fractionalized edge states
and non-local string order parameters. When perturbing away from the SU(2)
symmetric point, both the edge states and the string order can be
eliminated. Remarkably, the Haldane phase can still remain stable, given
that certain symmetries are preserved. I.e., the non-trivial topological
character of the Haldane phase is protected by symmetry, even though the
Haldane phase itself does not break any symmetry spontaneously.

We have shown that the non-trivial ``topological'' nature of the Haldane
phase of $S=1$ chains is reflected by a double degeneracy of the entire
entanglement spectrum. The degeneracy is protected by the same set of
symmetries which protects the Haldane phase, and cannot be lifted unless
either a phase boundary to another, ``topologically trivial'' phase is
crossed, or the symmetry is broken. The Haldane phase is protected by any of
the following symmetries: spatial inversion symmetry, time reversal symmetry
or the dihedral symmetry D$_2$ (rotations by $\pi$ about a pair of orthogonal axes). This result on the symmetry protection is completely consistent with
what was obtained from different arguments.\cite{Pollmann-2009a} The degeneracy of the
entanglement spectrum can be used to characterize the Haldane phase in the
most general situation, in which edge modes and string order may be absent.
(see TABLE~\ref{sec:char}).

The degeneracy of the entanglement spectrum in the Haldane phase is proven
by examining how the Schmidt eigenstates transform under a projective
representation of the symmetry group of the system. The transformation laws
contain phase factors, which are constrained to take discrete values by
symmetry. If these phase factors are non-trivial, they require a degeneracy
in the entanglement spectrum. Depending on which phase factors take
non-trivial values, several distinct ``Haldane-like'' states are possible.
This offers a scheme to classify all possible gapped phases of a
one-dimensional system, given its symmetry group. Such a general
classification will be the subject of a forthcoming paper.\cite{Turner-2009}
\begin{table}[htp]
\begin{center}
\begin{tabular}{l|c|c|c|}
symmetry & string order & edge states & degeneracy \\ \hline
D$_2$ (=Z$_2\times$Z$_2$) & yes & yes & yes \\
time reversal & no & yes & yes \\
inversion & no & no & yes \\
&  &  &
\end{tabular}%
\end{center}
\caption{The different symmetries which can stabilize the Haldane phase. For
each class of symmetries, the table shows whether string order, edge states,
or the degeneracy of the entanglement spectrum are necessarily present. The
symmetry under $\protect\pi $ rotations about a pair of orthogonal axes is
represented by the dihedral group D$_2$.}
\label{sec:char}
\end{table}

The degeneracy of the entanglement spectrum is a highly non-local property,
and is not easily related to physical observables. Nevertheless, the
degeneracy of the entanglement spectrum has direct physical consequences. It
means that in the Haldane phase, the entanglement of the system across any
cut cannot drop below the minimum value of $\ln(2)$. This can be observed,
for example, by adiabatically weakening a bond to zero. In the Haldane
phase, if inversion symmetry is preserved throughout this process, it leaves
the two disconnected halves of the system in a finitely entangled state. In
a ``topologically trivial'' state, on the other hand, the two halves can be
completely decoupled and form an unentangled product state after the process
has ended. The non-zero residual entanglement is reflected in correlation
functions of physical observables belonging to the two halves of the system.
Such an adiabatic weakening process could, at least in principle, serve as a
way to experimentally distinguish the Haldane phase from other,
``topologically trivial'' phases.

\section*{Acknowledgment}

We thank Ashvin Vishwanath, Ehud Altman, Emanuele Dalla Torre,
Joel E. Moore, Michael Levin, Masaaki Nakamura and Yasuhiro
Hatsugai for useful discussions.  E.~B. was supported by the NSF
under grant DMR-0757145. F.~P. and A.~M.~T. acknowledge support
from ARO grant W911NF-07-1-0576. M.~O. was supported by KAKENHI
grants No. 20654030 and 20102008. The authors acknowledge the IPAM
workshop QS2009 at which our work on the entanglement spectrum of
the Haldane phase was initiated.

\appendix
\section{Generalization of the symmetry-protected Haldane phase}

\label{sec:gen}

We have seen above that the Haldane phase of spin-1
antiferromagnetic chains can be characterized by a double
degeneracy of the the entanglement spectrum, which can be traced
back to the non-trivial transformation law of its Schmidt
eigenstates under certain symmetry operations. The double
degeneracy of the Haldane phase is protected either one of the
three symmetries  (dihedral group of $\pi$-rotations about two
orthogonal axes, time-reversal symmetry, or bond centered
inversion symmetry). In fact, there are various generalizations of
the Haldane phase, which are protected by modified symmetries.

For example, in magnetism, Dzyaloshinskii-Moriya (DM) interaction
generally arises if the system lacks inversion symmetry about the
center of bond. In the case of the one-dimensional chain, the DM
interaction is given as
\begin{equation}
 \sum_j \vec{D}_j \cdot (\vec{S}_j \times \vec{S}_{j+1}) .
\end{equation}
The following two cases often appear in models of magnetism: a
uniform DM interaction $\vec{D}_j = \vec{D}$ and a staggered DM
interaction $\vec{D}_j = (-1)^j \vec{D}$. Let us assume $\vec{D} =
(0,0,D)$ (parallel to $z$-axis).

The DM interaction, which is also known as antisymmetric exchange
interaction, clearly breaks inversion symmetry about the bond.
Here we consider the Hamiltonian of a $S=1$ chain
\begin{equation}
 H_{\mbox{\scriptsize DM}} = J \sum_j \vec{S}_j \cdot \vec{S}_{j+1}
 + B_z \sum_j S^z_j
+ \sum_j \eta^j
\vec{D} \cdot (\vec{S}_j \times \vec{S}_{j+1}),
\label{eq:HwithDM}
\end{equation}
where $\eta=1$ for the uniform and $\eta=-1$ for the staggered DM
interaction case. This model breaks all the three symmetries we
have discussed above, if $B_z, \vec{D} \neq 0$. Thus one might
expect that the Haldane phase is no longer well defined in this
model. However, it turns out that the double degeneracy of the
entanglement spectrum, and thus the well-defined Haldane phase,
survives for $D>0$. This can be simply understood because the
Hamiltonian can be transformed\cite{SEA-DM-1992,Derzhko1994} to
\begin{eqnarray}
 \tilde{H}_{\mbox{\scriptsize DM}} &=&
 U^{\dagger}_{G} H_{\mbox{\scriptsize DM}} U^{\vphantom{\dagger}}_{G} \nonumber \\
 &=& J \sum_j S^z_j S^z_{j+1} +
J_{\perp} \sum_j ( S^x_j S^x_{j+1} + S^y_j S^y_{j+1} )\nonumber\\
&+& B_z \sum_j S^z_j. \\
\label{DMtransf}
\end{eqnarray}
Here,  $J_{\perp} = \sqrt{J^2 + D^2}$, if we choose
\begin{equation}
 U_{G} = e^{i \sum_j j \alpha S^z_j},
\end{equation}
for the uniform DM interaction, and
\begin{equation}
 U_{G} = e^{i \sum_j (-1)^j (\alpha/2) S^z_j},
\label{eq:stagDMgauge}
\end{equation}
for the staggered DM interaction,
with $\alpha = \tan^{-1}{(D/J)}$.
The resulting Hamiltonian is simply the standard XXZ
antiferromagnetic chain in an magnetic field $B_z$.
The inversion symmetry of the model guarantees
a double degeneracy in the entanglement spectrum,
and hence protects the Haldane phase.

In the context of the original Hamiltonian, however,
the symmetry that protects the double degeneracy
is somewhat obscured.
The inversion $\mathcal{I}$ acts on the transformed Hamiltonian as
\begin{equation}
 \tilde{H}_{DM} \to \mathcal{I}^{\dagger} \tilde{H}_{DM} \mathcal{I} .
\end{equation}
Here we define the inversion $\mathcal{I}$ so that
site $j$ goes to site $1-j$.
We find that the modified symmetry of
the original Hamiltonian is the invariance under
\begin{equation}
 H_{DM} \to {\mathcal{I}'}^{\dagger} H_{DM} \mathcal{I}',
\end{equation}
where
\begin{equation}
 \mathcal{I}' = U^{\vphantom{\dagger}}_{G} \mathcal{I} U_{G}^{\dagger} =
  \left\{ \begin{array}{cc}
   e^{i \alpha \sum_j (2j-1) S^z_j} \mathcal{I} & \mbox{(uniform DM)} \\
   e^{i \alpha \sum_j (-1)^j S^z_j} \mathcal{I} & \mbox{(staggered DM)}
\end{array}
\right. .
\end{equation}
Namely, it is the invariance under inversion with an
appropriate ``twist'' (rotation of each spin about $z$-axis).

The invariance under $\mathcal{I}'$, which protects the Haldane
phase, is not a generic symmetry and may be broken rather easily
by perturbations which could occur naturally. For example, if the
uniform magnetic field were applied to $x$-direction instead of
$z$-direction in the Hamiltonian~(\ref{eq:HwithDM}), it is clear
that the model is no longer invariant under $\mathcal{I}'$.

For a staggered DM interaction, a uniform field in $x$-direction
leaves a staggered magnetic field $\propto \sum_j (-1)^j S^x_j$
after the transformation. The staggered field breaks inversion
symmetry about the bond center and thus eliminates the degeneracy
in the entanglement spectrum. That a staggered field destroys the
Haldane phase was noticed earlier.\cite{TsukanoNomura-JPSJ1998}

As another example of physical interest, let us discuss the
``Haldane-Insulator'' (HI) phase in the extended Bose-Hubbard
model (EBHM). We will show that the HI phase is protected by a
similar mechanism. The model Hamiltonian of the EBHM reads
\begin{eqnarray}
H_{BH}&=&-t\sum_{j}(b_{j}^{\dagger }b^{\vphantom{^{\dagger }}}_{j+1}+%
\mbox{H.c.})  \label{HBH} \\
&&+\frac{U}{2}\sum_{j}n_{j}\left( n_{j}-1\right)
+V\sum_{j}n_{j}n_{j+1} \nonumber
\end{eqnarray}%
where we assume a filling of one bosons per site
($\langle n\rangle =1$) and $t,U,V>0$.
In Ref.~\onlinecite{Berg-2008}, it has been shown that the EBHM
has a phase which is analogous to the Haldane phase. This phase
was termed a Haldane Insulator (HI).

The symmetries of the EBHM are translation, time-reversal, inversion,
and particle conservation. It is useful to consider an effective
spin-1 model by truncating the Hilbert space of each site to states with $%
n=0,1,2$, which is strictly justified in the large $U$ limit. This
modification is not expected to be important in this limit, since states
with $n>2$ are higher in energy. The corresponding effective pseudospin
Hamiltonian reads
\begin{eqnarray}
H_{\mathrm{eff}} &=&-t\sum_{j}\left( S_{j}^{+}S_{j+1}^{-}+\mbox{H.c.}\right)
+\frac{U}{2}\sum_{j}(S_{j}^{z})^{2}  \nonumber \\
&&+V\sum_{j}S_{j}^{z}S_{j+1}^{z}+H^{\prime }  \label{Heff}
\end{eqnarray}%
where we have introduced the pseudospin operator $S^{z}=n-1$, and $H^{\prime
}$ contains other terms which break the \textquotedblleft
particle-hole\textquotedblright\ symmetry of $H_{\mathrm{eff}}$, which is
represented in the pseudospin language by a $\pi $ rotation about the $x$
axis. This spurious symmetry is not crucial for the stability of the HI\
phase, as we shall show below.

Ignoring $H^{\prime }$, the Hamiltonian $H_{\mathrm{eff}}$ is very
similar to $H_{0}$ in Eq. (\ref{H0}), with the exception that the
$S_{j}^{+}S_{j+1}^{-}+\mbox{H.c.}$ (\textquotedblleft
hopping\textquotedblright ) term is of opposite sign. As a result,
the HI phase of (\ref{Heff}) is not protected by inversion
symmetry ($\mathcal{I}$). The phase $e^{i\phi _{\mathcal{I}}}$,
which is the parity of the ground state under inversion about a
bond, is equal to $+1$ in this case; the ground state of a system
with the ordinary, negative, sign for the kinetic energy cannot
have nodes.

However, the HI phase is protected by a modified symmetry instead.
The effective Hamiltonian can be mapped to the antiferromagnetic
spin Hamiltonian~(\ref{H0}) by a staggered rotation of spins by
$\pm \pi/2$ about $z$ axis, alternatingly on even and odd sites.
The staggered rotation is given by the unitary transformation of
the same form as eq.~(\ref{eq:stagDMgauge}), but now with
$\alpha=\pi$. We note that, if we increase the DM interaction from
zero to infinity for the antiferromagnetic chain, $|\alpha|$
changes from zero to $\pi/2$. Thus the present case is distinct
from the antiferromagnetic chain with DM interactions. The
transformation changes the sign of the hopping term to negative;
in the spin chain context this makes in-plane exchange interaction
antiferromagnetic as in Eq.~(\ref{H0}).

Following the discussion for a staggered DM interaction, and using
$\alpha=\pi$, we find that the HI phase is protected by invariance
under the operation
\begin{equation}
 \mathcal{I}' = e^{i \pi \sum_j S^z_j} \mathcal{I} .
\end{equation}
As an interesting example, a staggered field in $x$-direction is
now invariant under $\mathcal{I}'$ (and thus does not break the
double degeneracy of the entanglement spectrum), while a uniform
field in the same direction is not.

In terms of the discussion
in Sec.~\ref{sec:DGDG} , the HI phase is characterized by
$\phi _{\mathcal{I}^{\prime}}=\pi $ and $\phi _{\mathcal{I}}=0$,
and is thus distinct from the usual
Haldane phase of $H_{0}$, with $\phi _{\mathcal{I}^{\prime }}=0$
and $\phi _{\mathcal{I}}=\pi $.
This shows that these two states cannot be connected
adiabatically while preserving either $\mathcal{I}$ or
$\mathcal{I}^{\prime} $.

\section{Schmidt decomposition of a $\protect\phi _{\mathcal{I}}=\protect\pi
$ state}

Let us prove Eq. (\ref{Sch_asym}). We consider an inversion-symmetric MPS\ $%
\left\vert \Psi \right\rangle $ defined on a finite chain of an even length $%
2L$ and assume that $\left\vert \Psi \right\rangle $ is characterized by $\phi _{\mathcal{I}%
}=\pi $. The MPS is written as

\begin{eqnarray}
\left\vert \Psi \right\rangle &=&\sum_{\{m_{j}\}}V_{L}^{T}\Gamma
_{m_{1}}\Lambda \dots \Gamma _{m_{L}}\Lambda  \nonumber \\
&&\times \Gamma _{m_{L+1}}\Lambda \dots \Gamma _{m_{2L}}V_{R}|m_{1}\dots
m_{2L}\rangle \text{.}
\end{eqnarray}%
Since we are interested in
bulk properties in the limit $L\rightarrow \infty$, we assume a sufficiently long chain with position--
independent matrices $\Gamma_{m}$.
$V_{L}$ and $V_{R}$ are $\chi $
dimensional column vectors which define the boundary conditions (to
be specified later).  Describing the boundary
conditions in this way is possible as long as there are no edge modes,
which is generically true when only inversion symmetry is present
(Otherwise, the edges states of the two ends may require some extra care.). Since $\left\vert \Psi \right\rangle $ is
invariant under inversion, the matrices $\Gamma _{m}$ satisfy Eq.
(\ref{reflect}). Applying this relation to the matrices $\Gamma
_{m_{L+1}}\dots \Gamma _{m_{2L}}$, we get
\begin{eqnarray}
\left\vert \Psi \right\rangle &=&e^{-iL\theta
}\sum_{\{m_{j}\}}V_{L}^{T}\Gamma _{m_{1}}\Lambda \dots \Gamma _{m_{L}}\Lambda
\nonumber \\
&&\times U_{\mathcal{I}}\Gamma _{m_{L+1}}^{T}\Lambda \dots \Gamma
_{m_{2L}}^{T}U_{\mathcal{I}}^{\dagger }V_{R}|m_{1}\dots
m_{2L}\rangle \text{.}
\end{eqnarray}
Now, we choose boundary conditions such that
$V_{R}=U_{\mathcal{I}}V_{L}$. The wavefunction $|\Psi \rangle$ can
be written as
\begin{equation}
|\Psi \rangle =e^{-iL\theta }\sum_{\alpha ,\beta }\lambda _{\alpha }\left(
U_{\mathcal{I}}\right) _{\alpha \beta }|\alpha \rangle \overline{|\beta
\rangle }\text{,}
\end{equation}%
where
\begin{equation}
|\alpha \rangle =\sum_{\{m_{j}\}}\left( V_{L}^{T}\Gamma _{m_{1}}\Lambda
\dots \Gamma _{m_{L}}\right) _{\alpha }|m_{1}\dots m_{L}\rangle \text{,}
\end{equation}%
and $\overline{|\alpha \rangle }=\mathcal{I}|\alpha \rangle $. Since $\phi _{%
\mathcal{I}}=\pi $, the Schmidt eigenvalues $\lambda _{\alpha }$ are all
doubly degenerate (see Sec.~\ref{inv}). Let us order the $\lambda _{\alpha }$'s such that $%
\lambda _{2\alpha -1}=\lambda _{2\alpha }$ for every $1\leq \alpha
\leq \chi $. The matrix $U_{\alpha \beta }$ commutes with $\Lambda
$. Therefore, it must have a block-diagonal form with $2\times 2$
blocks on the diagonal. Since $U^{T}=-U$ (which follows from $\phi
_{\mathcal{I}}=\pi $, as shown in Sec. \ref{inv}), the blocks on
the diagonal of $U$ are all of the form $e^{i\eta _{\alpha
}}i\sigma _{2}$, where $\eta _{\alpha }$ is a phase. Therefore, we
write $|\Psi \rangle $ as
\begin{equation}
|\Psi \rangle =\sum_{\alpha =1}^{\chi /2}\lambda _{2\alpha -1}\left( |\alpha
,1\rangle \overline{|\alpha ,2\rangle }-|\alpha ,2\rangle \overline{|\alpha
,1\rangle }\right) \text{,}  \label{sch}
\end{equation}
where $|\alpha ,j\rangle \equiv e^{i\frac{\eta _{\alpha }-L\theta_{\mathcal{I}}}{2}%
}|2\alpha -1+j\rangle $ ($j=1,2$) and $\overline{|\alpha ,j\rangle }=%
\mathcal{I}|\alpha ,j\rangle $. In the limit $L\rightarrow \infty
$, the states $ |\alpha ,j\rangle $ become orthonormal [as can be
shown from the canonical conditions (\ref{canonical})], and
therefore Eq. (\ref{sch}) is the Schmidt decomposition of $|\Psi
\rangle $. This concludes our proof.


\end{document}